\documentstyle[prl,twocolumn,aps,graphics,epsfig]{revtex}

\begin{document}
    \draft
\input{epsf.tex}

\title{Atom laser dynamics}

\author{Nicholas Robins$^{1}$,  Craig Savage$^{2}$,
and Elena A. Ostrovskaya$^{1}$}

\address{$^{1}$ Optical Sciences Center, The Australian National
University, Canberra ACT 0200, Australia \\
$^{2}$ Department of Physics and Theoretical Physics, The 
Australian 
National University, Canberra ACT 0200, Australia}

\maketitle

\begin{abstract}
 An ideal atom laser would produce an atomic beam with highly stable
 flux and energy.  In practice the stability is likely to be limited
 by technical noise and nonlinear dynamical effects.  We investigate
 the dynamics of an atom laser using a comprehensive one dimensional,
 mean-field numerical model.  We fully model the output beam and
 experimentally important physics such as three-body recombination. 
 We find that at high pump rates the latter plays a role in
 suppressing the high frequency dynamics, which would otherwise limit 
 the stability of the output beam.
\end{abstract}


\narrowtext

Optical lasers have had an enormous impact on science and technology,
due to the intensity and coherence of the light they produce.  We
present a theoretical study of the output beam of the analogous matter
wave device, known as the ``atom laser''.  These have been demonstrated
in a number of laboratories, although so far only in an unpumped mode
\cite{mewes1,Hagley,bloch,bloch2,anderson,martin}.

Theoretical studies have highlighted both the similarities and
the differences between optical and atom lasers
\cite{wise1,wise2,holly,hope,kneer,naraschewski}.  The differences
arise because atoms are more complex than photons: they have mass,
giving a different free space dispersion relation, and they interact
with each other, producing strong nonlinearities.  The latter
generates complex dynamics, potentially complicating certain
experimental measurements,  of quantum noise for example.  Nevertheless,
the output beam dynamics might be useful, in itself, as a probe of the
excitations of the trapped condensate \cite{Japha,Choi}.

Fundamentally, it is the coherence properties, such as linewidth and
intensity correlations, that best capture the unique physics of lasers
\cite{mandel,SavageBallagh}.  Coherence is also important for many
practical applications, such as interferometry.  A classical nonlinear
model of the laser as a noise driven van der Pol oscillator shows that
the noise power, in both the phase and the amplitude, decreases in
inverse proportion to the laser power \cite{Armstrong}.

Schawlow and Townes showed that optical lasers' first order coherence,
or linewidth, is ultimately limited by spontaneous emission, and also
decreases in inverse proportion to the laser power \cite{Schawlow}. 
Analogous results have been derived for atom lasers \cite{Graham}.  In
practice, however, the Schawlow-Townes limit is not achieved, because
the linewidth is limited by technical noise and by
dynamical effects rather than by quantum noise.  For example
relaxation oscillations, due to the nonlinear interaction between the
inversion and the light, are the primary determinant of the low
frequency noise spectrum in many optical lasers \cite{Yariv}.

In this paper we focus on one aspect of the nonlinear dynamics of atom
lasers: the frequency spectrum of the atom laser output beam density. 
In particular, we investigate how the spectrum depends on the pumping
rate.  This spectrum describes the atom laser nonlinear dynamics. An 
experimentally measured spectrum would also include genuine ``noise'' due to 
physical processes that we do not model, such as fluctuations in 
the trapping potential.
This investigation is motivated by the 
inverse relation between noise power and pumping rate in the optical
laser. We choose to focus on the density because it
can be measured experimentally using sensitive optical
techniques such as phase modulation spectroscopy \cite{Lye}.

A conclusion of our work is that three-body recombination plays a
major role in the high frequency dynamics.  As the pumping rate is
increased the spectral peaks weaken and move to higher frequencies. 
This knowledge will facilitate atom laser applications, and
measurements of their quantum noise.

\begin{figure}
\setlength{\epsfxsize}{6.0cm}
\centerline{\epsfbox{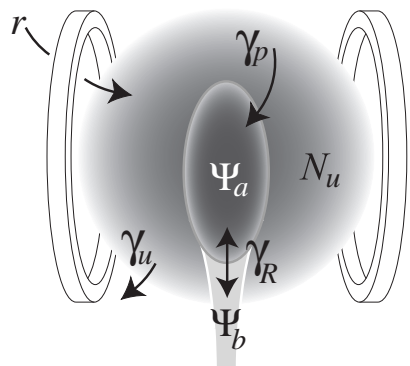}}
\caption[]{Schematic representation of our atom laser model.  The
rings represent coils producing the trapping potential.  The symbols
are defined in the text.  Atoms are injected into the non-condensed
fraction.  They can then either be lost or pump the
condensate.  They are coupled out of the condensate by a Raman
transition and then fall in the gravitational field.}
\label{fig1}
\end{figure}
Our atom laser model is schematically illustrated in Fig.~1.  It
consists of a dilute gas Bose-Einstein condensate trapped by a highly
anisotropic, effectively one-dimensional, potential.  This is pumped
from a reservoir of non-condensed trapped atoms.  The output atom
laser beam is generated by a Raman transition to an untrapped atomic
state.  Our model builds on that of Kneer {\em et al.}\cite{kneer} by
adding three-body recombination and a propagating output beam.  In the
following we will refer to the trapped Bose-Einstein condensate as the
``condensate'', and to the output beam as the ``beam''.  We make the
mean-field approximation in which the condensate and beam are
described by the classical fields $\Psi_a(x,t)$ and $\Psi_b(x,t)$
respectively, which obey Gross-Pitaevskii type equations
\cite{dalfovo}.  Our model is more complete than previous mean-field
treatments \cite{kneer,edwards,Schneider1,ballagh,band} since we
include pumping, output coupling, three-body recombination, and
explicitly model the dynamics of the pump reservoir, the condensate,
and the beam.  Although the extension to three dimensions is
straightforward, we do not report it here.  To describe the pumping we
use a phenomenological model, which mimics the pumping of an optical
laser \cite{kneer}.  It only depends on the total number of atoms in
the uncondensed component, $N_u(t)$, not on its spatial structure. 
The condensate atoms are coupled, by a reversible Raman transition
\cite{Hagley,edwards,moys}, to an untrapped electronic state forming
the atom laser beam.  The Raman transition imparts a momentum kick
$\hbar k$ to the out-coupled atoms.  The beam evolves under the
influences of gravity and atom-atom interactions, which are dominated
by those due to the condensate.  We have only considered the
particular case for which the Raman coupling is tuned to transfer
atoms starting from the center of the harmonic trap \cite{edwards}. 
Hence the output beam overlaps with the lower half of the condensate. 
Experimentally, three-body recombination is well described as a loss
process occurring at a rate proportional to the cube of the local
density \cite{burt97}.  We follow Kagan {\it et al.} \cite{Kagan98} in
incorporating three-body recombination into the atom laser
Gross-Pitaevskii equations.

Our model is defined by the following dimensionless equations:
\begin{equation}
\label{eq_main}
\begin{array}{l} 
{\displaystyle
\frac{d N_{u}}{dt}=r-\gamma_{u}N_{u}-\gamma_{p}N_{u}N_{a}},\\[9pt]
{\displaystyle
i\frac{\partial \Psi_{a}}{\partial t}=-\frac{1}{2}\frac{\partial^2 
\Psi_{a}}{\partial
x^2}+\frac{1}{2}x^{2}\Psi_{a}+U_{a}\Psi_{a}|\Psi_{a}|^2}\\[9pt]
{\displaystyle
 \qquad \quad +U_{ab}\Psi_{a}|\Psi_{b}|^2
 -i\gamma_{r}\Psi_{a} ( |\Psi_{a}|^4 +|\Psi_{b}|^4 )
 }\\[9pt]
{\displaystyle
 \qquad \quad 
 +\gamma_{R} e^{ikx}
\Psi_{b}+\frac{i}{2}\gamma_{p}N_{u}\Psi_{a}},\\[9pt]
{\displaystyle
i\frac{\partial \Psi_{b}}{\partial t}=-\frac{1}{2}\frac{\partial^2 
\Psi_{b}}{\partial x^2}+Gx\Psi_{b}+U_{b}\Psi_{b}|\Psi_{b}|^2}\\[9pt]
{\displaystyle
 \qquad \quad 
+U_{ab}\Psi_{b}|\Psi_{a}|^2
-i\gamma_{r}\Psi_{b} ( |\Psi_{a}|^4 +|\Psi_{b}|^4 )}\\[9pt]
{\displaystyle
 \qquad \quad 
 +\gamma_{R} e^{-ikx}\Psi_{a}.}
\end{array} 
\end{equation}
The model is made dimensionless using the characteristic trap length
$l=(\hbar/\omega m)^{1/2}$ and angular frequency $\omega$, with $m$
the atomic mass.  Hence time $t$, position $x$, and the fields are
measured in units of $\omega^{-1}$, $l$ and $l^{-1/2}$,
respectively.  Experimentally reasonable values of the parameters are:
a trap frequency $\omega \approx 125$ Hz \cite{mewes}, and the atomic
mass of sodium $m=3.8 \times 10^{-26}$ kg.  These give a time scale 
of $\omega^{-1}=8$ ms, and a length scale
of $l=4.7$ $\mu$m.

Atoms are injected into the uncondensed fraction at the rate $r$, are
lost at the rate $\gamma_{u}N_{u}$, and pump the condensate at the
rate $\gamma_{p}N_{u}N_{a}$, where
$N_{a}=\int^{\infty}_{-\infty}|\Psi_{a}|^2 dx$ is the total condensate
population.  $U_{a}$ and $U_{b}$ are the intra- and $U_{ab}$ the
inter- species two-body interaction coefficients.  The dimensionally
correct coefficients can be written as $U_{a,b,ab}=4\pi
a_{s}/(\hbar/\omega m)^{1/2}$ with $a_{s}$ the appropriate s-wave
scattering lengths \cite{edwards}.  We have previously found that
three-body recombination, with coefficient $\gamma_{r}$, is necessary
for the system to reach a quasi-stationary-state on a timescale
comparable to experimental condensate lifetimes \cite{us}.  The
three-body recombination rate $\gamma_{r} = 10^{-7}$ is chosen to
produce $N_a \approx 3500-9500$ condensed atoms at steady state, as
the pump rate varies from $r=200$ to $1600$.  For simplicity we assume
that the inter- and intra-species three-body recombination rates are
the same.  $G$ is the scaled gravitational acceleration \cite{G
footnote}.  Gravity has not been explicitly included for the trapped
atoms, since it is equivalent to a spatial shift of the trap
potential.  The Raman coefficient $\gamma_{R}$ is a function of the
Raman laser amplitude and detuning \cite{moys}.

Our numerical method is a split-step Fourier method \cite{taha} with a
fourth order Runge-Kutta in place of the usual nonlinear step.  We
have absorbing boundaries at the ends of the spatial grids to prevent
reflections.
\begin{figure}
\setlength{\epsfxsize}{7.0cm}
\centerline{\epsfbox{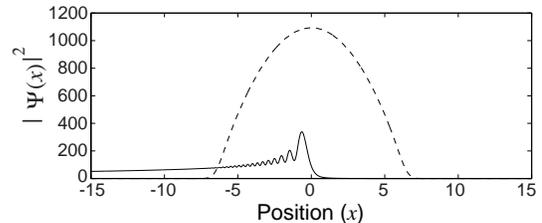}}
\caption[]{Densities of the condensate (dashed) and beam
(solid) versus position.  All quantities are dimensionless.  The
plotted beam density is 100 times the true value.  These results are a
numerical solution of Eqs.~(\ref{eq_main}) at $t=200$ 
after starting from a seed condensate.  Parameters are: $r=1600$,
$U_{a}=U_{b}=0.02$, $U_{ab}=0.01$, $G=12$, $\gamma_R=0.5$,
$\gamma_u=\gamma_p=0.1$, $\gamma_r=10^{-7}$ and $k=5$.}
\label{compare}
\end{figure}

Typical results for quasi-stationary-state spatial profiles of the
condensate and beam densities are shown in Fig.~2.  For our
parameters, the condensate shape is well described in the Thomas-Fermi
approximation.  Note that the plotted output beam density has been scaled up
by a factor of 100.  The flux in the beam is about 90 atoms per unit
dimensionless time, or about $1.1 \times 10^{4}$ atoms per second with
$\omega = 125$ Hz.  The prominent spatial oscillations of the
output beam, occurring within the extent of the condensate, have been
observed in previous work \cite{Schneider1,Schneider} and are related
to the oscillatory Airy type eigenfunctions of particles in the linear
gravitational potential.
\begin{figure}
\setlength{\epsfxsize}{7.0cm}
\centerline{\epsfbox{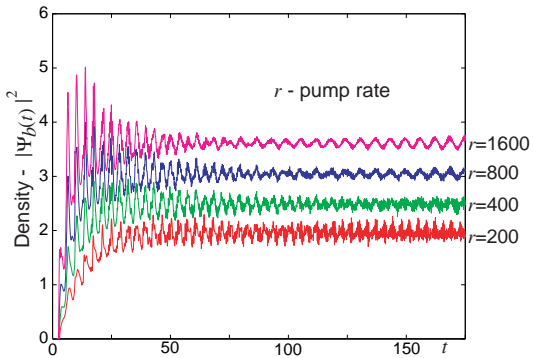}}
 \caption[]{The output beam density, $|\Psi_{b}(x_{0},t)|^{2}$, as a
 function of time at position, $x_{0} = -45$, for 
 different pump rates: from bottom to top $r=$ 200, 400, 800, and
 1600.  All quantities are dimensionless.  Other parameters are as for
 Fig.~\ref{compare}.}
\label{int1}
\end{figure}

In Fig.~\ref{int1} we show the output beam density
$|\Psi_{b}(x_{0},t)|^{2}$ as a function of time at a fixed position
far from the trapped condensate.  The time span has been chosen to be
longer than that of the transient dynamics associated with the growth
of the condensate from a seed, and short enough to be
experimentally accessible.  Quasi-stationary, nearly periodic dynamics
develop, becoming approximately harmonic for large pump rates $r$. We 
have verified that the quasi-stationary dynamics is independent of the 
particular form of the seed used to initiate the condensate growth.
The fundamental oscillation of the beam, at higher pump rates, arises
from the well known Kohn mode \cite{dalfovo}, corresponding to a rigid
``sloshing'' back and forth of the condensate at the trap frequency.
\begin{figure}
\setlength{\epsfxsize}{7.0cm}
\centerline{\epsfbox{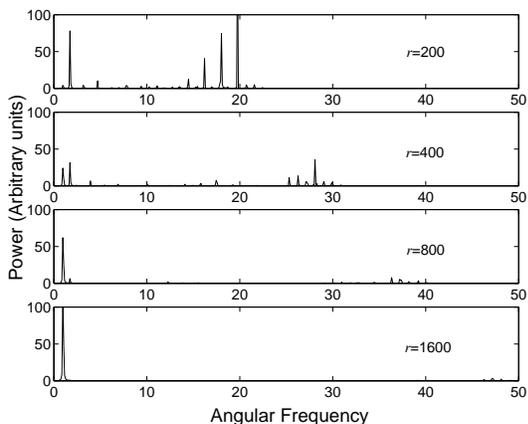}}
\caption[]{Frequency spectra of the time dependent density of the
fields in Fig.~\ref{int1}.  All quantities are dimensionless.  The
spectra are those of the time series from $t=112.5$ to $t=175$.  The pump
rates from top to bottom are $r=$ 200, 400, 800, and 1600.  The
normalization has been chosen so that the largest peak
height (that near 20 frequency units for $r$=200) is 100 units.  The
peaks just below 50 frequency units, for $r=1600$, are almost invisible
on this scale.}
\label{pumping}
\end{figure}

Fig.~\ref{pumping} shows the frequency spectra of the density time
series in Fig.~\ref{int1}, and is the central result of this paper. 
The spectra are those of the final, quasi-steady, parts of the time
series.  The frequencies are well approximated by the
eigen-frequencies of an uncoupled 1D condensate.  Kneer {\em et al.}
\cite{kneer} found these to be $\Omega=\omega \sqrt{{n(n+1)}/{2}}$
where $n=1,2,3,\ldots$, with the corresponding spatial modes given by
the Legendre polynomials $P_{n}(x)$.  The Kohn sloshing mode at
$\omega$ is $n=1$, and $n=2$ is the breathing mode at $\sqrt{3} \,
\omega$.  In this mode the width of the condensate oscillates, with a
corresponding density oscillation.  

Fig.~\ref{pumping} shows that as the pumping rate $r$ increases, the
slow beam oscillations change from the frequency of the condensate
breathing mode to that of the sloshing mode.  This is because the
condensate itself changes from breathing to sloshing at higher pumping
rates.  As Kneer {\it et al.} \cite{kneer} noted, the sloshing mode is
spatially asymmetric and hence must be excited by a spatially
asymmetric perturbation, such as the output beam.  This is consistent
with the observed increase of the sloshing mode power with the density
of the output beam, that is with pumping rate.  On the other hand,
mode damping due to three-body recombination increases rapidly with
density.  We have verified that in a simple model of the Kneer {\it et
al.} \cite{kneer} form, the breathing mode decays with pumping rate.

Although the frequencies in the beam are primarily determined by the
condensate, there is a complex relationship between the power spectra
of the condensate and of the beam.  In particular, the beam spectrum
reflects the spatial dependence of the condensate spectrum.  Consider,
for example, the condensate Kohn sloshing mode.  The magnitude of the
local density change varies with the spatial derivative of the density
profile.  It is maximum where the condensate has greatest slope, and
zero at the center where the slope is zero.  That the Kohn mode
frequency is prominent in the power spectrum of the output beam
indicates that the coupling between the condensate and the beam is not
confined to the center of the condensate.  Similar observations apply for
the high frequency components of the power spectrum, the envelopes of
which differ in detail between the condensate and the beam.

As the pumping rate increases, Fig.~4 shows that the frequency of the
high frequency group increases and its power decreases rather
dramatically.  A decrease of spectral power with pumping rate, due to
gain saturation, is characteristic of optical lasers. 
Since three-body recombination contributes to gain saturation in atom
lasers, we have investigated its role in a
simplified pumped atom laser model similar to that of Kneer {\it et
al.} \cite{kneer}.  That is, with a spatially uniform phenomenological
loss rather than an explicitly modelled output beam.  Without
three-body recombination the condensate density power spectrum simply
rolls off at high frequencies.  Adding three-body recombination
produces a high frequency spectral group, such as seen in Fig.~4. 
Hence it is an important factor in the dynamics of the system.

In conclusion, we have explored some of the rich and complex 
behavior of the atom laser with a model incorporating important
experimentally relevant physics, such as three-body recombination. 
The latter was found to strongly influence the spectrum of the
atom laser output beam.  This work is a step towards the future goal
of understanding and measuring the quantum noise properties of the
atom laser.


\begin{references}
 
\bibitem{mewes1} M.-O. Mewes, M. R. Andrews, D. M. Kurn, D. S. 
Durfee, C. G. Townsend, and W. Ketterle, \prl {\bf 78}, 
582 (1997).

\bibitem{Hagley} E. W. Hagley, L. Deng, M. Kozuma, J. Wen, K. 
Helmerson, S. L. Rolston and W. D. Phillips,
Science {\bf 283}, 1706 (1999).

\bibitem{bloch}I. Bloch, T. W. Hansch, and T. Esslinger,
\prl {\bf 82}, 3008 (1999). 

\bibitem{bloch2}I. Bloch, T. W. Hansch, and T. Esslinger,
Nature {\bf 403}, 166 (2000).

\bibitem{anderson} B.P Anderson and M.A. Kasevich, Science {\bf 282}, 
1686 (1998).

\bibitem{martin} J.L. Martin {\it et al.}, J. Phys. B: At. Mol. Opt. 
Phys. {\bf 32}, 3065 (1999).

\bibitem{wise1}H. Wiseman and M. Collett, \pra {\bf 202}, 246 (1995).

\bibitem{wise2} H. Wiseman, A. Martins and D. Walls, Quant. 
Semiclass.  Opt.  {\bf 8}, 737 (1996).

\bibitem{holly} M. Holland, K. Burnett, C. Gardiner, J.I Cirac and P. Zoller,
\pra {\bf 54}, R1757 (1996).

\bibitem{hope} J. J. Hope, G. M. Moy, M. J. Collett and C. M. 
Savage, \pra {\bf 61}, 023603 (2000).

\bibitem{kneer} B. Kneer, T. Wong, K. Vogel, W. P. Shleich, and D. F. 
Walls, \pra {\bf 58}, 4841 (1998).

\bibitem{naraschewski} M. Naraschewski, A. Schenzle, and H. Wallis, 
\pra {\bf 56}, 603 (1997).

\bibitem{Japha} Y. Japha, S. Choi, K. Burnett, and  Y. Band, 
\prl {\bf 82}, 1079 (1999).

\bibitem{Choi} S. Choi, Y. Japha, and K. Burnett,
\pra {\bf 61}, 063606 (2000).

\bibitem{mandel} L. Mandel and E. Wolf. 
{\em Optical coherence and quantum optics.}
(Cambridge University Press: 1995).

\bibitem{SavageBallagh} R.J. Ballagh and C.M. Savage, {\it
Bose-Einstein Condensation: from atomic physics to quantum fluids. 
Proceedings of the 13th Physics Summer School}, edited by C.M. Savage
and M. Das (World Scientific, Singapore, 2000).

\bibitem{Armstrong} J.A. Armstrong and A.W. Smith, Prog.  Opt. {\bf VI},
211 (1967).

\bibitem{Schawlow}  A.L. Schawlow and C.H. Townes, 
Phys. Rev. \textbf{112}, 1940 (1958).

\bibitem{Graham} R. Graham, Phys. Rev. Lett. {\bf 81}, 5262 (1998).

\bibitem{Yariv} A. Yariv, {\it Quantum Electronics} (Wiley, NY, 1989).

\bibitem{Lye} J.E. Lye, B.D. Cuthbertson, H.-A. Bachor, J.D. Close,  
J. Opt. B: Quant. Semiclass. Opt. {\bf 1}, 402 (1999).
 
\bibitem{dalfovo} F. Dalfovo, S. Giorgini, L. P. Pitaevskii and S. 
Stringari, Rev. Mod. Phys. {\bf 71}, 463 (1999).

\bibitem{edwards} M. Edwards, D. A. Griggs, P. L. Holman, 
C. W. Clark, S. L. Rolston and W. D. Phillips, J. Phys. B: At. Mol. 
Opt. Phys. {\bf 32}, 2935 (1999).

\bibitem{Schneider1} J. Schneider and A. Schenzle,
App. Phys. B {\bf 69}, 353 (1999).

\bibitem{ballagh} R. J. Ballagh, K. Burnett, and T. F. Scott,  
Phys. Rev. Lett. {\bf 78}, 1607 (1997).

\bibitem{band} Y. B. Band, P. S. Julienne and M. Trippenbach, \pra
{\bf 59}, 3823 (1999).

\bibitem{moys} G. M. Moy, J. J. Hope and C. M. Savage, \pra 
{\bf 55}, 3631 (1997).

\bibitem{burt97} E.A. Burt, R.W. Ghrist, C.J. Myatt, M.J. Holland, E.A. 
Cornell, and C.E. Wieman, \prl {\bf 79}, 337 (1997).

\bibitem{Kagan98} Yu. Kagan, A. E. Muryshev,  G. V. Shlyapnikov,
Phys. Rev. Lett. {\bf 81}, 933 (1998)

\bibitem{mewes} M.-O. Mewes, M. R. Andrews, N. J. van Druten, D. M. 
Kurn, D. S. Durfee, and W. Ketterle, Phys. Rev. Lett. {\bf 77}, 416 
(1996).

\bibitem{us} N. Robins, C. Savage and E. Ostrovskaya, {\it Walls Memorial
Volume}, edited  by H.J. Carmichael {\em et  al.}, (Springer 2000).

\bibitem{G footnote} For reasons of computational efficiency, we found
it convenient to use $G=12$.  This is because higher values produce
faster atoms with shorter de Broglie wavelengths, requiring smaller
spatial grid sizes, and hence bigger computations.  For the preceding
parameters $g=9.8$ ms$^{-2}$ corresponds to $G=132$.  Physically, low
$G$ values might correspond to the atom laser beam propagating in a
tilted wave-guide.  Alternatively, a trap frequency about five times
higher ($\omega \approx 625$ Hz) would make $G=12$ correspond to
$g=9.8$ ms$^{-2}$.

\bibitem{taha} T. R. Taha and M. J. Ablowitz, J. Comp.  Phys.  {\bf
55}, 203 (1984).

\bibitem{Schneider} J. Schneider and A. Schenzle,
\pra {\bf 61}, 053611 (2000).


\end{references}
\end{document}